\documentstyle[12pt,epsf,epsfig,wrapfig]{article}
\newcommand{\be}{\begin{equation}}

\newcommand{\ee}{\end{equation}}

\newcommand{\bea}{\begin{eqnarray}}

\newcommand{\eea}{\end{eqnarray}}

\newcommand{\kt}{\vec{k}}
\newcommand{\p}{\vec{p}}

\begin{document}
\begin{center}
{\bfseries PROBING TRANSVERSITY IN ELECTROPRODUCTION OF TWO VECTOR MESONS. }
\vskip 5mm
D.Yu.~Ivanov${}^{1,2}$,
{ B.~Pire}${}^{3}$,
\underline {L.~Szymanowski}${}^{4}$ and
{ O.V.~Teryaev}${}^{5}$
\vskip 5mm
{\small
(1) {\it
 Institut f{\"u}r Theoretische Physik, Universit{\"a}t
   Regensburg,  D-93040 Regensburg, Germany
}
\\
(2) {\it
Institute of Mathematics, 630090 Novosibirsk, Russia
}
\\

(3) {\it
CPhT, {\'E}cole Polytechnique, F-91128 Palaiseau, France
}
\\
(4) {\it
  So{\l}tan Institute for Nuclear Studies,
Ho\.za 69, 00-681 Warsaw, Poland
}
\\
(5) {\it
Bogoliubov Lab. of Theoretical Physics, JINR, 141980 Dubna, Russia
}
\\
}
\end{center}
\vskip 5mm

\begin{abstract}
Electroproduction of two vector mesons
with a large rapidity gap between them  provides the first feasible
 selective access to chiral-odd GPDs \cite{Ivanov:2002jj}.
 The  scattering amplitude of the process
$ \gamma ^* N \to \rho_L^0 \rho_T^+ N'$
 may be represented
as a convolution
of an impact factor describing the $\gamma ^* \to \rho_L^0$ transition
and an amplitude describing the $N\to \rho_T^+ N'$ transition, analogously to deeply exclusive
electroproduction of a meson, the
virtual photon being replaced by a Pomeron. 
The production of a transversely polarized vector meson $\rho_T^+$
selects a chiral-odd GPD in the proton.
\end{abstract}

\vskip 8mm

The study of transversity \cite{trans} is of fundamental interest for understanding the
 spin structure of nucleons. Generalized Parton Distributions (GPDs) are the non-perturbative
objects encoding the information about the quark and gluon proton
structure in the most complete way \cite{GPD}. While the chiral even GPD
may be probed in various hard exclusive processes, no single one has yet
be proven to be sensitive to  transverse spin dependent chiral-odd GPDs \cite{COLLINS,COGPD}.

In the massless quark limit, the chiral-odd functions may appear
only in pairs in a non-vanishing scattering amplitude, so that chirality
flip encoded in one of them is compensated by another. The natural probe
for the forward chiral-odd distributions is the Drell-Yan process,
containing the convolution of chiral-odd distributions of quark and
antiquark \cite{tra}.
Its nonforward analog is provided by the hard exclusive
production of a transversely polarized vector meson, where the
quark transversity distribution in the nucleon  is
substituted by the chiral-odd GPD. The generalized (skewed) 
transversity distribution in nucleon target 
$F^{T}_\zeta(x)$ described by the
polarization vector $n^\mu$  is defined by the formula
\bea
\label{FT}
&&\langle N(p_{2'},n)|\bar q(0) \sigma^{\mu\,\nu} q(y)| N(p_2,n) \rangle
= \\
&&=\bar u(p_{2'},n)\sigma^{\mu \,\nu} u(p_2,n)
\,\int\limits_0^1\,d\,x_1 \;\left[
e^{-ix_1(p_2y)}
F^{T\;q}_\zeta(x_1) - e^{ix_2(p_2y)} F^{T\;\bar q}_\zeta(x_1)\right]\;,
\nonumber
\eea
where $\sigma^{\mu\,\nu}=i/2[\gamma^\mu\,,\gamma^\nu], \;x_2=x_1 - \zeta$.

However, the simplest realization of this idea, namely the hard
exclusive electroproduction of a transversely polarized vector meson,
results in a zero contribution \cite{MPW}, \cite{DGP}. We suggest another process
which  allows to access the chiral-odd GPD's by "substituting" to the
virtual photon a hard two gluon exchange, {\it i.e.}  a perturbative
Pomeron (${\cal P}$) in the lowest order,
coming from a virtual photon $\to $ meson transition.
 Let us consider the specific process
\be
\label{2mesongen}
\gamma^*_L (q)\;\; N (p_2) \to  \rho_{L}^0¥(q_\rho)\;\; \rho^+_{T}¥(p_\rho)
N'(p_{2'})\;,
\ee
shown in Fig. 1 of scattering of a  virtual  photon on a
nucleon $N$, which leads via two gluon exchange to the production
of  two vector mesons separated by a large
rapidity gap  and the scattered nucleon $N'$. We choose a charged vector meson $\rho^+$ to
 select quark antiquark exchange with the nucleon line.
We consider the kinematical region where the rapidity gap 
between $\rho^+$ and
$N'$ is much smaller than the one between
$\rho^0$ and $\rho^+$, that is the energy of the system 
($\rho^+\; -\; N'$) is smaller
than the energy of the system ($\rho^0 -\; \rho^+$) but still large enough to
justify our approach  (in particular much larger than baryonic resonance masses). 

Let us first summarize  the kinematics of the process.
 The momenta are parametrized in terms of two light-like Sudakov vectors
$p_{1/2}$ as follows~:
\bea
&& q^\mu = p_1^\mu -
\frac{Q^2}{s}p_2^\mu\;,\;\;\;\;q^2=-Q^2\,,\;\;\;\;s=2(p_1p_2)\nonumber \\
&& q_\rho^\mu = \alpha p_1^\mu +
\frac{\p^{\;2}}{\alpha s}p_2^\mu + p_\perp^\mu \;,\;\;\;\;\;p_\perp^2 =
-\p^{\;2}
\nonumber \\
&& p_\rho^\mu = \bar \alpha p_1^\mu + \frac{\p^{\;2}}{\bar \alpha s}p_2^\mu
- p_\perp^\mu \;,\;\;\;\;\;\bar \alpha \equiv 1-\alpha \nonumber \\
&& p_{2'}^\mu = p_2^\mu (1- \zeta)
\label{Sud}
\eea
where 
$\zeta$ is the skewedness parameter  which can be written in
terms of
the two meson invariant mass
\be
s_1 = (q_\rho + p_\rho)^2 = \frac{\p^{\;2}}{\alpha \bar \alpha}
\label{s1}
\ee
and the photon virtuality $Q^2$ as
\be
\zeta = \frac{1}{s}\,\left(Q^2 + s_1  \right)\,.
\label{zeta}
\ee
The $\rho^+(p_\rho)-$meson - target invariant mass equals
\be
s_2 = (p_\rho + p_{2'})^2 = s\, \bar \alpha\,\left(1-\zeta  \right)\,.
\label{s2}
\ee
The kinematical limit with a large rapidity gap between the two mesons in the
final state is
obtained by demanding that $s_1$ is very large, being of the order of $s$
\be
s_1 =s\, \zeta\,,\;\;\;\;s_1 \gg Q^2,\,\,\p^{\;2}\,,
\label{s1gap}
\ee
whereas $s_2$ is kept constant but  large enough to justify the use
of  perturbation theory in the
collinear subprocess ${\cal P} N \to \rho^+_T N'$ and the application of
the GPD framework
\be
\label{s2gap}
s_2 \to \frac{\p^{\;2}}{\zeta}\,\left(1-\zeta  \right) = constant\,.
\ee
In terms of the longitudinal fraction $\alpha$ the limit
with a large rapidity gap corresponds
to taking the limits
\be
\label{alphagap}
\alpha \to 1\,,\;\;\;\;\;\bar \alpha s_1 \to \p^{\;2}\,,\;\;\;\;\;\;\zeta
\sim
1\,.
\ee

We have shown \cite{Ivanov:2002jj} that in  such kinematical
circumstance, the Born term for this process is calculable consistently within
the collinear factorization method. The final result is represented as an
integral (over  the longitudinal momentum fractions of the quarks)  of
the product of two amplitudes: the first one describing
the transition $\gamma^* \to \rho_L^0$ via two gluon exchange and
the second one  describing the subprocess
${\cal P}\;N\;\to \;\rho^+\;N'$ which is
closely related to the electroproduction process 
$\gamma^*\,N \to\rho^+\,N'$
where  collinear factorization theorems allow to separate  the long distance dynamics  expressed
through the GPDs from a perturbatively calculable coefficient function. The case of transversally
 polarized vector meson $\rho_{T}^+$  involves the chiral-odd GPD. The hard scale
appearing in this process  is supplied by the relatively large  momentum transfer
$p^2$ in the two gluon channel, i.e. by the virtuality of the Pomeron.

\begin{wrapfigure}{R}{8cm}
\mbox{\epsfig{figure=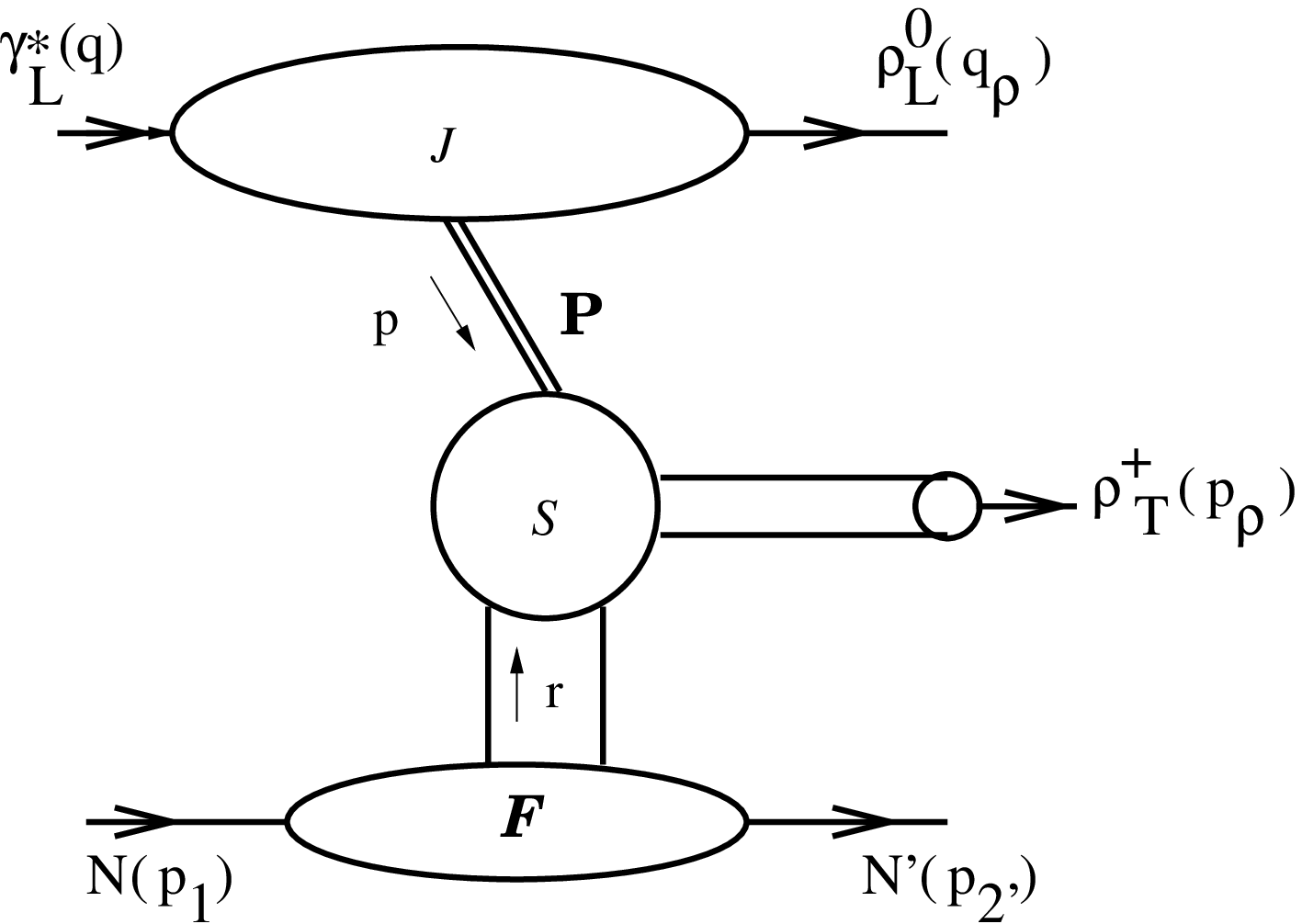,width=7.8cm,height=6cm}}
{\small{\bf Figure 1.} Factorization of the process $\gamma^*_L (q)\;\; N (p_2) \to  \rho_{1}¥(q_\rho)\;\; \rho^+_{2}¥(p_\rho)
N'(p_{2'})\;$ in the asymmetric
kinematics discussed in the text. ${\bf P}$ is the hard Pomeron. }
\end{wrapfigure}

Such a process is a representative of a new class of hard
reactions whose QCD description within the collinear factorization scheme
involves in the described above kinematics
the impact factor $J$ appearing naturally in Regge-type perturbative description
based on the BFKL evolution and the collinear distributions whose evolutions are governed by DGLAP-ERBL
equations.

We have choosen the kinematics so that the nucleon gets no transverse
momentum in the process. Let us however note that in principle one may allow a finite
momentum transfer, small with respect to $|\p|$. This case will involve
additional GPDs in the expressions to follow.

We have shown that the collinear factorization holds
at least in the Born approximation. 
The resulting scattering amplitude has a very compact form and it reads : 

\bea
\label{CON}
&&{\cal M}^{\gamma^*\,p\,\to \rho_L^0\, \rho^+_T\,n}=
 -i\,\sin \theta \;8\pi^2 \zeta s \alpha_s f_T \sqrt{1-\zeta}
\frac{C_F}{N\,(\p^{\;2})^2}
\nonumber \\
&&\int\limits_0^1
\frac{\;du\;\phi_\perp(u)}{ \,u^2 \bar u^2 }
 J^{\gamma^*_L \to \rho^0_L}(u\p,\bar u\p)
 \left[ F_\zeta^{T\;u}(u\zeta)-F_\zeta^{T\;\bar d}(u\zeta)\right]\;,
\eea
where  $\theta$ is the angle between the transverse polarization vector of
the target
$\vec{n}$ and the polarization vector
$\vec{\epsilon}_T$ of the produced $\rho^+_T-$meson.

$J^{\gamma^*_L \to \rho^0_L}$ is the  impact factor 
\be
\label{ifgamma}
J^{\gamma^*_L \to \rho^0_L}(\kt_1,\kt_2)=  -   f_\rho \frac{e
\alpha_s 2\pi
 Q}{N_c\sqrt{2}} \int\limits_0^1 dz\;z\bar z \phi_{||}(z)P(\kt_1,\kt_2)\;,
\ee
with
\bea
\label{P}
P(\kt_1,\kt_2=\p-\kt_1)=&&\frac{1}{z^2\p^{\;2}+m_q^2 +Q^2z\bar z} +
\frac{1}{{\bar z}^2\p^{\;2}+m_q^2 +Q^2z\bar z} \nonumber \\
&&-\frac{1}{(\kt_1-z\p\,)^2+m_q^2 +Q^2z\bar z} -
\frac{1}{(\kt_1-\bar z \p\,)^2+m_q^2 +Q^2z\bar z}\;.
\eea

The transversely polarized $\rho$-meson is described by means of its chiral-odd
light-cone distribution amplitude \cite{BalBr} defined
by the matrix element
\be
\label{rho+T}
\langle \rho_T(p_\rho,T) \mid \bar q(x) \sigma^{\mu \nu} q(-x)\mid 0
\rangle =i f_T \left(p_\rho^{\mu}\epsilon^{*\nu}_T -
p_\rho^{\nu}\epsilon^{*\mu}_T
\right)
\int\limits_0^1 du e^{-i(2u-1)(p_\rho x)}\;\phi_\perp(u)\;,
\ee
where  $\phi_\perp(u)=6u\bar u$ and $f_T(\mu)=160\pm10\,$MeV at the
scale
$\mu=1\,$GeV.

Another example is the process  with $\rho^0$ replaced by
heavy $J/\Psi-$meson. One could study either $\gamma^*_L \;\; N  \to
J/\Psi_L\;\; \rho_T^+\;\;   N'$ involving the impact factor
\bea
\label{charmLL}
&&J^{\gamma^*_L \to
J/\Psi_L}(\kt_1,\kt_2=\p-\kt_1)=
\\
&&=-\frac{8\pi\,e\,\alpha_s\,Q\,f_{J/\Psi}}{3\,N}\;
\left(\frac{1}{\p^{\;2}+Q^2+4\,m_c^2} - \frac{1}{(2\,\kt_1 -\p)^2 +Q^2+
4\,m_c^2}
\right) \;, \nonumber
\eea
or the process $\gamma^*_T \;\; N  \to
J/\Psi_T\;\; \rho_T^+\;\;   N'$ for which the impact factor $J^{\gamma^*_T \to
J/\Psi_T}$ has the form
\bea
\label{charmTT}
&&J^{\gamma^*_T \to
J/\Psi_T}(\kt_1,\kt_2=\p-\kt_1)=
\\
&&=\frac{4\,e\,\alpha_s\,\pi\,m_c\,f_{J/\Psi}}{3\,N}\;(\vec \varepsilon
\,\vec \epsilon^{\;*})\,\left( \frac{1}{\p^{\;2} + Q^2 +4\,m_c^2} -
\frac{1}{(2\,\kt_1-\p)^2 + Q^2 +4\,m_c^2}   \right)\;.
>\nonumber
\eea
These impact factors are obtained from the corresponding ones for light
quarks by applying a standard non-relativistic approximation for
the $J/\Psi$-meson vertex, i.e.
by approximating
the distribution amplitude  by
$\phi_{j/\Psi}(z)=\delta(z-1/2)$.
The coupling constant $f_{J/\Psi}^2 =\frac{27\,m_{J/\Psi}\,\Gamma_{J/\Psi
\to e^+
e^-}}{16\,\pi\,\alpha_{em}^2}$ is expressed in terms of the
width $\Gamma_{J/\Psi \to e^+\,e^-}$
and
$\alpha_{em}={e^2}/{4\pi}$.
This last process (\ref{charmTT}) can be of course also studied both for
virtual as well as
real photon.

One can also study the transversity in the inelastic DVCS: $\gamma^*_{L/T}
\;\; N  \to \gamma\;\; \rho_T^+\;\;   N'$. The expressions for the
corresponding impact factors
$J^{\gamma^*_T \to \gamma_{T'}}$ and
$J^{\gamma^*_L \to \gamma_{T'}}$
  are also known and can be found e.g. in
\cite{BalKu} (
Eq.~(11)  and Eq.~(12), respectively).

 The case with $Q^2=0$, i.e. the photoproduction
at large momentum transfer is more complicated and may require to take into
account both the perturbative and the hadronic (non-perturbative)
contributions (see e.g. \cite{larget}).

In conclusion, the chiral-odd GPD may now be accessed in a feasible reaction, namely
$\gamma^*\,p\;\to \;\rho^0_L\,\rho^+_T\,n$ which  can be described consistently 
within the collinear factorization
approach in the specific kinematics where the two mesons are produced with a large
rapidity gap.  Higher order studies are
necessary  to  establish the validity of this factorization property beyond the Born order. 
An estimate of the cross-section of this reaction requires a knowledge of the
chiral-odd GPD. No model has yet been proposed for this quantity.
Further improvement of the theoretical understanding of hadronic impact factors may help us
 to access this chiral-odd GPD also in hadronic diffractive reactions at RHIC and LHC.

\vskip.2in
{\Large \bf Acknowledgments}
\vskip.2in
The work of B.P. and L.Sz. is supported by the french-polish scientific 
agreement Polonium.
D.I. acknowledges the support of the Alexander von Humboldt Foundation
and RFBR 02-02-17884. O.T. was supported by RFBR Grant 03-02-16816.

\vskip.2in
Q. (A. Sandacz, SINS, Warsaw):\\
Is it possible to consider $\rho^0$-meson instead of $\rho^+$ at the
rapidity close to the nucleon rapidity?

\vskip.1in
A. Yes, it is possible to replace a $\rho^+$-meson by a $\rho^0$-meson in the
rapidity region of the target nucleon. At the Born order which we have
 discussed the
scattering amplitude has a very similar form as for the $\rho^+$-meson case.  
\end{document}